\documentclass{article}

\usepackage{arxiv}

\usepackage[utf8]{inputenc} 
\usepackage[T1]{fontenc}    
\usepackage{url}            
\usepackage{booktabs}       
\usepackage{amsfonts}       
\usepackage{nicefrac}       
\usepackage{microtype}      
\usepackage{lipsum}		
\usepackage{graphicx}
\usepackage{natbib}
\usepackage{doi}

\usepackage{algorithmic}
\usepackage{amsmath,amssymb,amsfonts}
\usepackage{graphicx}
\usepackage{color}
\usepackage{comment}
\usepackage{enumitem}
\usepackage{hyperref}
\usepackage{listings}
\usepackage{textcomp}
\usepackage[table,xcdraw]{xcolor}
\usepackage{tabularx}

\lstdefinelanguage{sparql}{
    tabsize=1,
    keywordstyle=\color{magenta},
	morestring=[b][\color{blue}]\",
	morekeywords={SELECT,CONSTRUCT,DESCRIBE,ASK,WHERE,FROM,NAMED,PREFIX,BASE,OPTIONAL,FILTER,GRAPH,LIMIT,OFFSET,SERVICE,UNION,EXISTS,NOT,BINDINGS,MINUS,a,BIND,COUNT,GROUP,ORDER,BY,DESC,AS,DISTINCT},
	sensitive=true,
	breakatwhitespace=false,         
    breaklines=true,                 
    captionpos=b,                    
    keepspaces=true,                 
    numbers=left,                    
    numbersep=4pt,                  
    showspaces=false,                
    showstringspaces=false,
    showtabs=false,
    basicstyle=\scriptsize,
    literate={\ \ \ \ }{{\ }}1,
    xleftmargin=0.5cm 
}

\title{An Ontological Approach to Analysing Social Service Provisioning}

\author{ \href{https://orcid.org/orcid=0000-0001-7444-6310}{\includegraphics[scale=0.06]{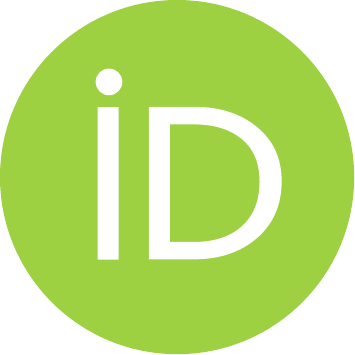}\hspace{1mm}Mark S. Fox}\thanks{Supported, in part, by HelpSeeker Technologies Inc. and the Canadian Digital Technology Supercluster Fund.} \\
	Centre for Social Services Engineering\\
	University of Toronto\\
	Toronto, Canada\\
	\texttt{msf@mie.utoronto.ca} \\
\And
\href{https://orcid.org/0000-0001-6201-8781}{\includegraphics[scale=0.06]{orcid.pdf}\hspace{1mm}Bart Gajderowicz} \\
	Centre for Social Services Engineering\\
	University of Toronto\\
	Toronto, Canada\\
\texttt{bartg@mie.utoronto.ca} \\
\And
\href{https://orcid.org/0000-0002-5877-9681}{\includegraphics[scale=0.06]{orcid.pdf}\hspace{1mm}Daniela Rosu} \\
	Centre for Social Services Engineering\\
	University of Toronto\\
	Toronto, Canada\\
\texttt{drosu@mie.utoronto.ca} \\
\And
Alina Turner \\
HelpSeeker Technologies Inc.\\
Calgary, Canada \\
\texttt{alina@helpseeker.org}\\
\And
Lester Lyu \\
	Centre for Social Services Engineering\\
	University of Toronto\\
	Toronto, Canada\\
\texttt{lvds2000@gmail.com} \\
}

\renewcommand{\headeright}{Fox, Gajderowicz  et~al.}
\renewcommand{\shorttitle}{Ontology for Analysing Social Service Provisioning}

\hypersetup{
pdftitle={An Ontological Approach to Analysing Social Service Provisioning},
pdfauthor={Mark S. Fox, Bart Gajderowicz, Daniela Rosu, Alina Turner, Lester Lyu},
pdfkeywords={ontology, semantics and reasoning, social services, decision support system, data model},
}

\begin{document}
\maketitle


\begin{abstract}
This paper introduces ontological concepts required to evaluate and manage the coverage of social services in a Smart City context. Here, we focus on the perspective of key stakeholders, namely social purpose organizations and the clients they serve. The Compass ontology presented here extends the Common Impact Data Standard by introducing new concepts related to key dimensions: the \textit{who} (Stakeholder), the \textit{what} (Need, Need Satisfier, Outcome), the \textit{how} (Service, Event), and the \textit{contributions} (tracking resources). The paper first introduces key stakeholders, services, outcomes,  events, needs and need satisfiers, along with their definitions. Second, a subset of competency questions are presented to illustrate the types of questions key stakeholders have posed. Third, the extension's ability to answer questions is evaluated by presenting SPARQL queries executed on a Compass-based knowledge graph and analysing their results. 

\end{abstract}

\section{Introduction}
Within any region, there exists a myriad of social services designed to meet citizens' needs. The provisioning of these services is provided by government agencies at the national, provincial/state and local levels. They are also provided by Non-Government Organizations such as non-profits, charities and social enterprises. The number of agencies involved is surprisingly large. For example, Canada has over 170,000 charities and non-profits, and the United States has over 1.54 million. The providers of social services distinguish themselves along many dimensions, two of which are the type of service they provide and the target group (beneficiary stakeholder) the service is designed to help 


Services range across many areas, such as housing, food, and training. Stakeholders are differentiated across dimensions such as age, gender and medical conditions. The variety of services differentiated by target stakeholders leads to endless variations of what is 
determining whether the plurality of social services  are meeting their citizens' needs. 

As our society evolves into a richer mosaic of cultures, ethnicities, etc., cities are challenged in determining whether needs are being met, by various stakeholder distinctions. The broad categories used for social service provisioning are no longer sufficient (\cite{DAquin2015,Daga2016,Lytras2018}). A deeper representation of stakeholder characteristics, their needs and how social services satisfy them is required. This paper defines such a representation: the Compass ontology.

In section \ref{sec:usage-cqs}, we present a set of coverage questions posed by social services stakeholders. In sections \ref{sec:client}, \ref{sec:service}, \ref{sec:needs}, and \ref{sec:event}, we identify key ontological components required to answer the proposed coverage questions and describe the Compass ontology\footnote{The Compass extension is prefixed with the \textit{cp:} namespace. When no namespace is provided, \textit{cp:} is implied.}, an extension to the Common Impact Data Standard (CIDS)\footnote{The CIDS ontology 
is prefixed with the \textit{cids:} namespace.} (\cite{Fox2021}). Finally, in section 
\ref{sec:eval} we evaluate the extension by showing sample SPARQL queries and results generated from a knowledge graph built using the Compass extension\footnote{Code available at \url{https://github.com/csse-uoft/ieee-isc2-2022}.}. In section \ref{sec:conclusion} we conclude by discussing the evaluation results and discuss the state of the ontology and future work.

\begin{figure*}[ht]
    \centerline{\includegraphics[width=0.7\textwidth]{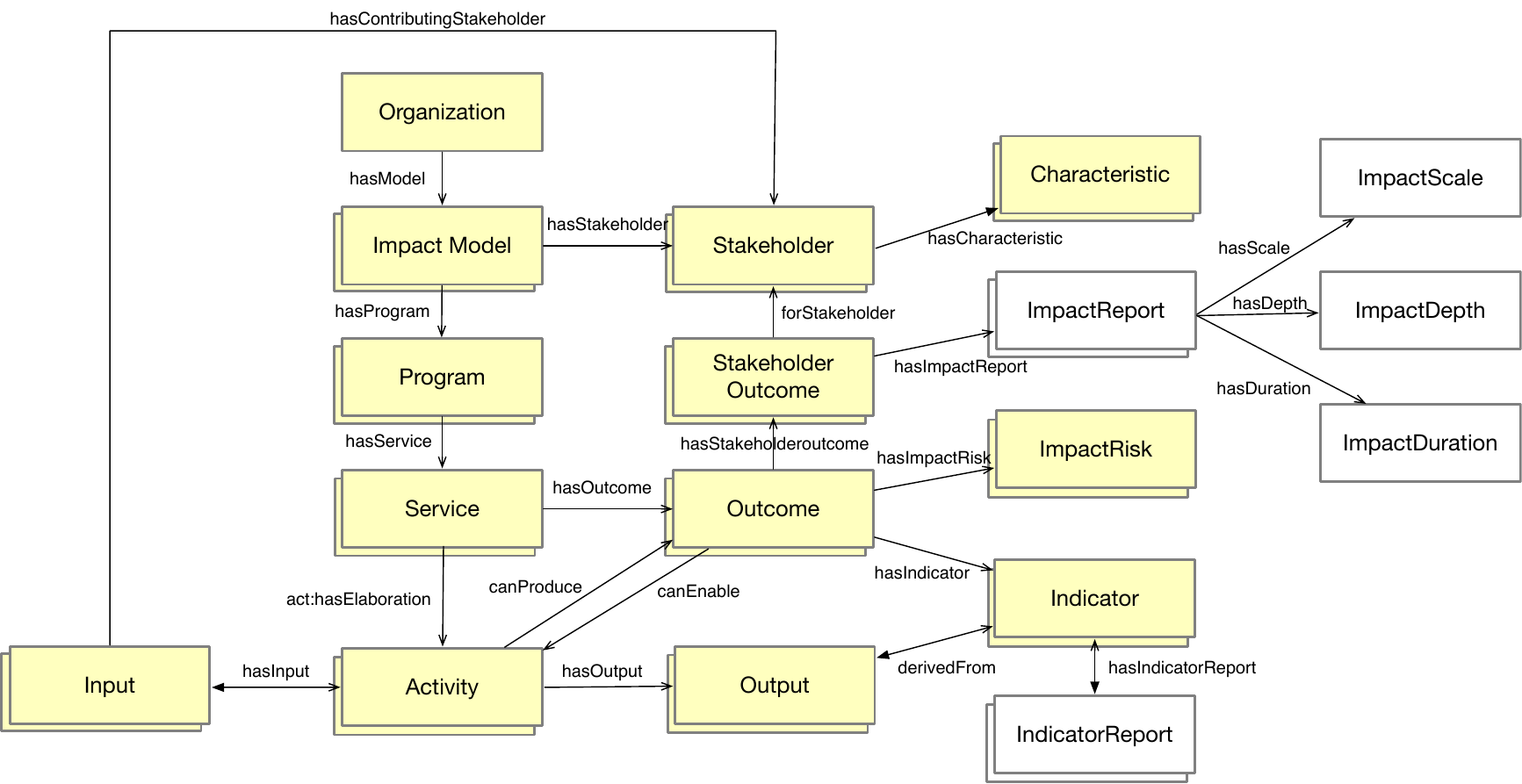}}
    \caption{Common Impact Data Standard Core Classes.}
    \label{fig:cids}
\end{figure*}

\section{Usage Scenarios and Competency Questions}
\label{sec:usage-cqs}

\subsection{Modeling Service Coverage}


Coverage of service refers to the variety of services available in a particular location and how accessible those services are to the communities that need them. Hence, to determine service coverage, we must identify several key ontological components. The Compass ontology extends the Common Impact Data Standard depicted in Figure \ref{fig:cids}. CIDS defines one or more Impact Models for an Organization. Each Impact Model can have one or more Programs, each having set of general Stakeholders, Outcomes and Indicators. Each Program has one or more Services with more specific Stakeholders, Outcomes and Indicators. Services are composed of Activities that have Inputs, Outputs and associated Indicators.

The \textit{cids:Outcome} class defines the resulting state of service provisioning, generally in a larger scope that captures the long-term impact of a service on a community. The \textit{cids:ContributingStakeholder} class represent stakeholders that perform some activities towards achieving the outcome by providing resources (e.g. funding) or administering services (e.g. front-line workers). The \textit{cids:BeneficialStakeholder} class represents stakeholders impacted by the outcome, whether in a positive, negative, or neutral way. These include clients that benefit directly from a service as well as organizations that receive resources or share activities with other contributing stakeholders. 

Services, clients and needs may be categorized along a variety dimensions that external organizations define themselves or are defined differently in related sectors. Hence, the Compass ontology includes a set of taxonomies that define unique codes for each dimension that reference external representations.

\label{subsec:usage-scenarios}

\begin{figure*}[ht]
    \centerline{\includegraphics[width=0.8\textwidth]{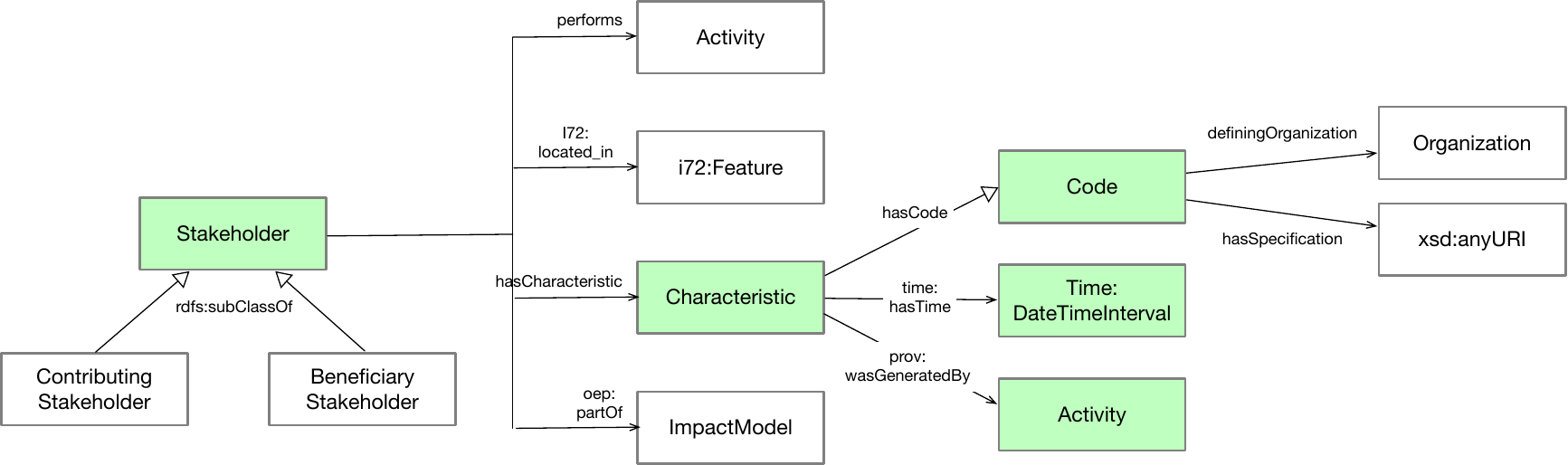}}
    \caption{Common Impact Data Standard \textit{Stakeholder} class and its core classes.}
    \label{fig:stakeholder}
\end{figure*}


\subsection{Client Stakeholders Usage Scenario and CQs}
\label{subsec:client-intro}

Clients can be identified by their level of need. \textbf{Low need clients} are those that are generally able to care for themselves, but require some assistance. 
\textbf{High need clients} have similar needs, but require additional assistance, such as advocacy on behalf of a renter to negotiate rent or assistance for people with a history of trauma in performing daily tasks, such as shopping or filling out forms with sensitive information. \textbf{Super High Need Clients} require assistance with the majority if not all of their daily tasks and access to services. Assistance may include gathering information about available food programs, access to governmental benefits and pension, and help resolving barriers to service. The questions clients ask include:
\begin{description}
    \item[Client CQ-1] (1.1) What services am I eligible for and (1.2) what barriers exist for me to access them?
    
 

    \item[Client CQ-2] What shelter services are available in my community?

    \item[Client CQ-3] Which services match my needs?

    \item[Client CQ-4] I have a history of trauma; am I going to have to talk about that in the program?
 

    \item[Client CQ-5] (5.1) What are the rules at the shelters? (5.2) Do I have to be sober? (5.3) Where is the building? 
   
    \item[Client CQ-6] If I don't like the program what are my other options?
    
    
    
    \item[Client CQ-7]  (7.1) What are you going to do with my data in the program? (7.2) Can the police or government access the system?

\end{description} 

\subsection{Service Stakeholders Usage Scenario and CQs}
\label{subsec:service-intro}

Service stakeholders are those that work for social purpose organizations to deliver services to clients. These include partner organizations that support each other as contributing stakeholders towards a common outcome. 

We will focus on three key service stakeholders selected by subject matter experts. A \textbf{complex needs coordinator} is a person who manages clients with complex needs. They are responsible for managing client needs and ensuring services are available when needed. A \textbf{service manager} is responsible for managing individuals and resources who deliver services, such as food inventory as well as peer counsellors, shelter facility maintenance staff, and volunteers.  Finally, \textbf{front-line service worker} is a role that involves interacting directly with clients on a daily basis. They are responsible for being familiar with the needs of the client, their current status in a program, and the people most likely to help or hinder the client's well being. The questions service providers may ask include:
\begin{description}
    \item[Service CQ-1]  What are the key demographics of the highest systems users (age, ethnicity, immigration, etc)? 

    \item[Service CQ-2] How long did client \#2 stay in counselling?

    \item[Service CQ-3] What is causing barriers to accessing Housing?


    \item[Service CQ-4] What other programs or services could I refer clients to at this time? 
    
    

\end{description}

\subsection{Outcome-Related Stakeholders Usage Scenario and CQs}
\label{subsec:outcome-intro}

Outcome related stakeholders are those that focus on ``big-picture`` questions asking \textit{how} communities are impacted by specific outcomes, and \textit{how} resources (i.e. need satisfiers) are distributed. In some instances, the outcomes may be related to a specific need, a specific demographic group, or a combination of the two, such as ``addiction services for people experiencing homelessness'' or ``the long-term housing needs of marginalised communities (LGBTQ, new immigrants, etc)''. Depending on their role, such stakeholders may or may not be part of the organizations that directly deliver services to clients. For example, an \textbf{executive director} is responsible for the growth of a service providing organization and ensuring that key performance indicators are met. They focus on meeting the targets set out by the organization and reaching out to community and service partners. On the other hand, \textbf{senior managers at funding agencies} are \textit{contributing stakeholders} interested in how their funding is used to achieve their target outcome. The questions such stakeholders may ask include:

\begin{description}
        \item[Outcome CQ-1] What are the associated priority demographic groups?


    \item[Outcome CQ-2] What is the representation of these groups in my community?


    \item[Outcome CQ-3] What housing services are available by gender, racialized identity, age etc.?

    \item[Outcome CQ-4] What areas is funding currently going toward and how much of this funding is my organization getting? 

    \item[Outcome CQ-5] What are our gaps and duplicates for services or funding?

\end{description}

\section{Client Pattern}
\label{sec:client}

The Compass \textit{Client} class connects to the main \textit{Service} and \textit{ClientNeed} class extensions through the \textit{cids:Characteristic}, \textit{cids:Stakeholder}, and \textit{cids:Code} classes. The \textit{Client} inherits all properties of \textit{5087-2:Person} (\cite{5087-2}) and satisfies needs of a \textit{cids:Stakeholder}. Figure \ref{fig:stakeholder} depicts main \textit{cids:Stakeholder} classes and properties, with further details in Fox et~al.(\cite{Fox2021}). 

Some properties of the \textit{Client} classes are ``coded properties'', meaning their range values are constrained by client taxonomy codes\footnote{Each code is a subclass of \textit{cids:Code}, and related to other classes through the \textit{cids:hasCode} property.  For example, \textbf{client taxonomy codes} are subclasses of the \textit{ClientCode} class, which is itself a subclass of \textit{cids:Code}. Each client code is an instance of a specific \textit{ClientCode} subclass. For example, the class \textit{CL-Age} can have several age categories, including \textit{cp:INST-Toddler}, \textit{cp:INST-Young} and \textit{cp:INST-Adult}.  Similarly, \textit{ServiceCode} and \textit{NeedCode} define codes for services and needs.}. Different subcategories for basic demographics include Age, Ethnicity, Family status, Gender, Religion, and Sexuality. The taxonomy also lists different categories for a client's status along dimensions of Citizenship,  Finance, Education, Employment, and a Legal context.  It also provides codes for categories of a client's living situation, including shelter type, level of homelessness, and safety ranking. A client's constraints cover physical and mental health status as well as any non-health related constraints such as cultural barriers and social limitations. Finally, any property, coded or not, can be qualified with a \textit{Rank}  defined in the client taxonomy, such as acute, complex, severe or stable, as well as a \textit{Temporality}, such as past, long-term,  and ongoing. \textit{Client} properties related to needs and outcomes are defined in section \ref{sec:needs}.

\begin{figure}[ht]
    \centerline{\includegraphics[width=0.6\textwidth]{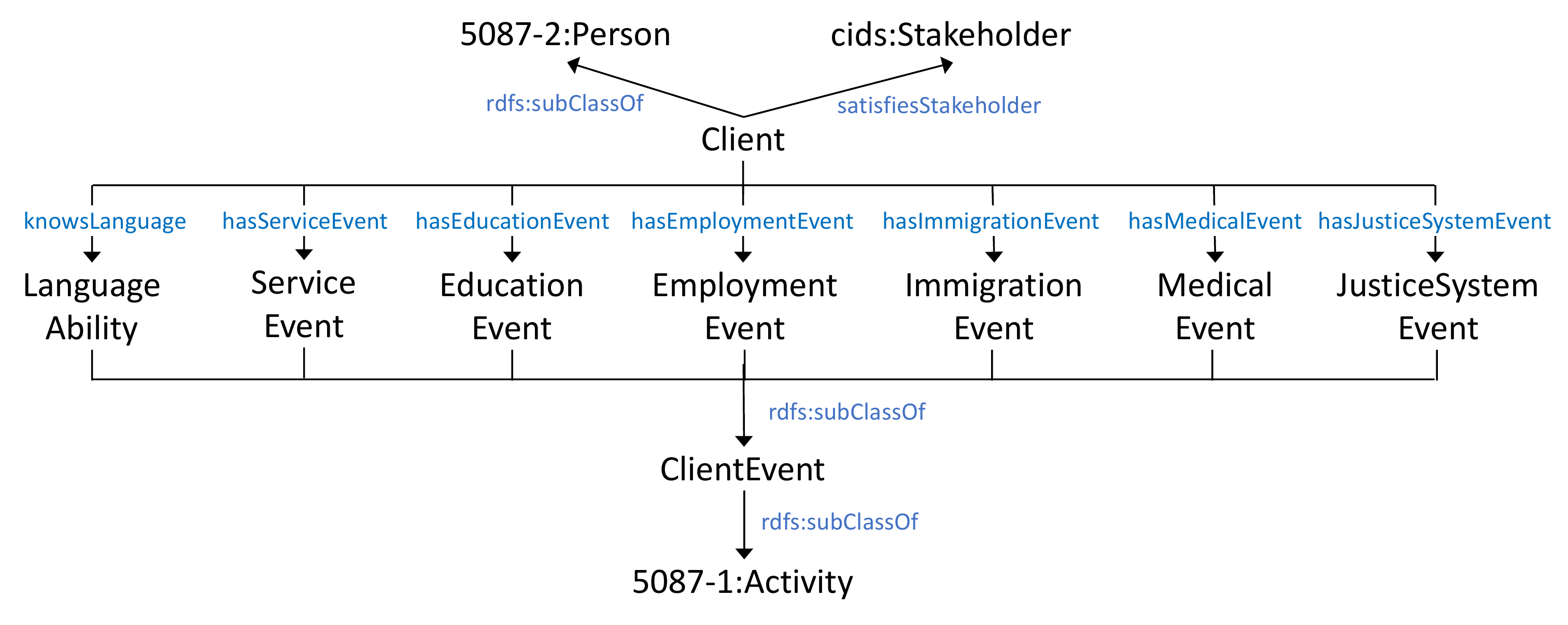}}
    \caption{Client class extension and related Event classes.}
    \label{fig:client_event}
\end{figure}

The Compass \textit{Client} extends the \textit{5078-2:Person} class with additional properties, as shown in Figure \ref{fig:client_event}. \textit{satisfiesStakeholder} is the stakeholder this client represents. The \textit{cids:Stakeholder} class provides the \textit{hasOutcome} property that identifies instances of \textit{StakeholderOutcome}, specifying outcomes experienced by the client. \textit{hasGender} is a coded property specifying the client's gender. \textit{hasEthnicity} is a coded property specifying known ethnicities of the client. \textit{memberOfAboriginalGroup} is a coded property identifying the aboriginal group the client is a member of, if any. \textit{hasReligion} is a coded property specifying known religions of the client. \textit{hasDependent} is a set of \textit{5087-2:Person} instances that identify the client's dependents, such as children or parents. \textit{schema:knowsLanguage} links to instances of \textit{LanguageAbility} that specifies languages known and their proficiency. 

\section{Service Pattern}
\label{sec:service}

\begin{figure*}[ht]
    \centerline{\includegraphics[width=0.8\textwidth]{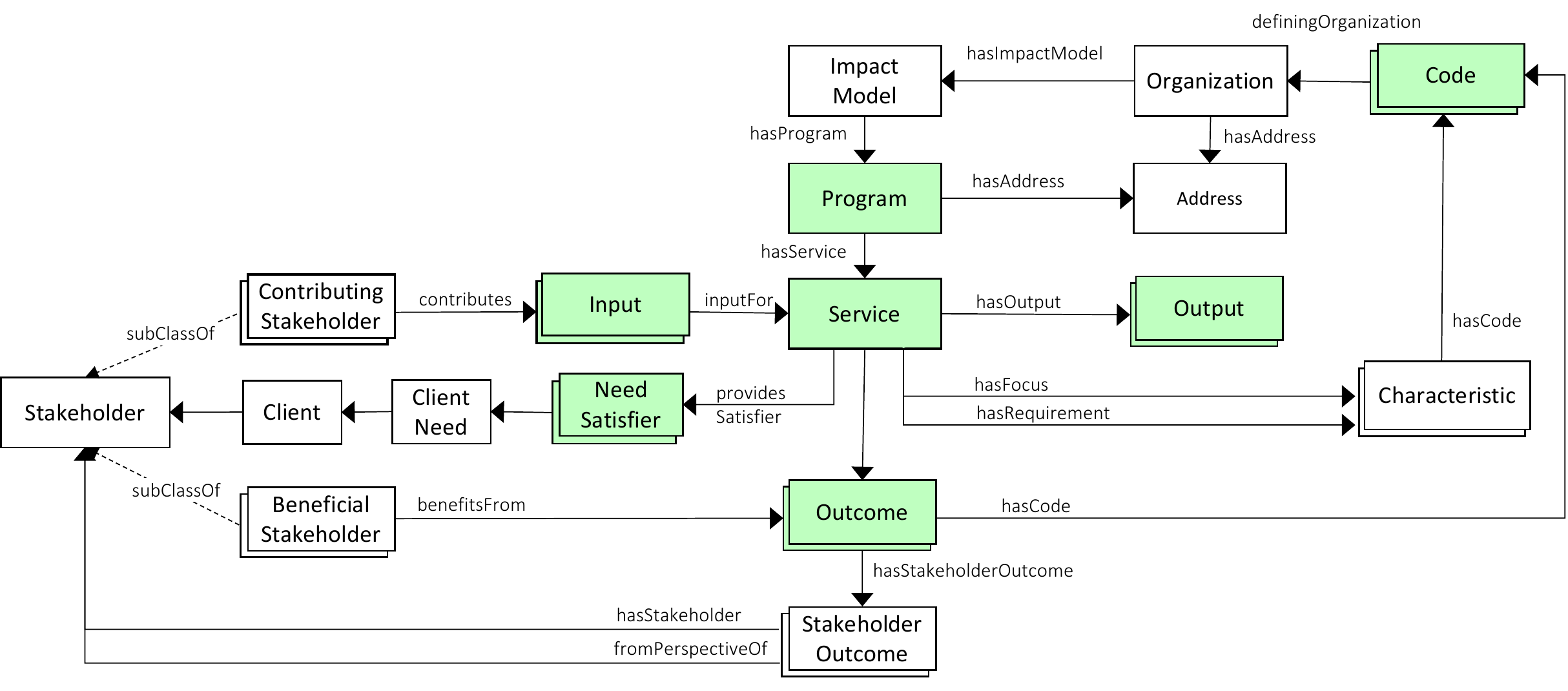}}
    \caption{Compass extension of the \textit{Service} class and its core classes.}
    \label{fig:service}
\end{figure*}

The \textit{cids:Service} class defines a set of activities, inputs, outputs, and outcomes that can be administered through a logic model and program. A service also defines a set of stakeholders that may contribute to or benefit from the service. The Compass \textit{Service} class extends \textit{cids:Service} and inherits all its properties. The core classes are shown in Figure \ref{fig:service}. The Compass \textit{Service} extension connects the main \textit{Client} and \textit{ClientNeed} classes through the \textit{cids:Characteristic}, \textit{cids:Stakeholder}, and \textit{ServiceCode} classes. Since a service may be administered by providers at different locations, the Compass \textit{Program} class extension introduces the \textit{ic:hasAddress} property that identifies the unique address for a program. 

The \textit{Service} class extension has the following properties. \textit{hasRequirement} identifies characteristics that limit who can use the service, listed in the \textbf{client taxonomy code} list. \textit{providesService} identifies codes for categories of services provided, as listed in \textbf{service taxonomy code list}, where each code is a subclass of \textit{ServiceCode}. For example, different types of education services are instances of the class \textit{CL-Education}, and include \textit{INST-Gradeschool}, \textit{INST-Highschool} as well as \textit{1\_on\_1\_Coaching} and \textit{INST-Educational\_Workshops}. A summary of services covered by the \textbf{service code taxonomy} is provided at the end of this section.

Next, the \textit{hasFocus} property identifies client characteristics that the service focuses on, listed in the client taxonomy code list. \textit{hasMode} specifies the mode with which the service is delivered, and can be one of \{\textit{in-person, phone, online, offline}\}. Finally, \textit{providesSatisfier} identifies the need satisfier this service provides, described section \ref{sec:needs}.

To define capacity of a service provisioning, we first need to define the community component of a service provider, meaning, the communities services target, if any. \textit{5078-2:CityDivision} class is extended from the ISO 5078-2 standard (\cite{5087-2}) to capture this functionality. A \textit{Community} class identifies two or more individuals that share some characteristics. The \textit{hasCommunityCharacteristic} property identifies characteristics of the community this class represents. The \textit{CommunityCharacteristic} class references the identifying characteristics that several individuals have in common that in turn make them a community. \textit{hasCommunityCharacteristic} identifies the characteristic that defines this \textit{Community}. \textit{hasNumber} is the number of people in the community limited to those that match these characteristic.

Clients can interact with service providers through a computer application, defined as \textit{Application}. The \textit{Application} class extends \textit{Service} class with an additional property, \textit{hasSource}, that identifies the unique resource locator (URL) or unique address where the application can be referenced.

The \textbf{service code taxonomy} includes categories of services as instances of the \textit{ServiceCode} class. For example, \textit{Cl-Health} includes short- and long-term medical services, including hospital stays and pharmaceutical services. \textit{CL-Cost} specifies the cost category of the service, including ``paid,'' ``sliding scale'' or ``through a subsidy.''  \textit{CL-Personal} services are those that clients require to live their daily lives, such as transportation and laundry services. \textit{Shelter} services provide housing programs for clients, including short- and long-term housing and specialized housing. Additional codes are defined for \textit{CL-Advocacy, CL-Referral, CL-Education, CL-Employment, CL-Relationships, CL-Finance, CL-Goods}, and \textit{CL-Food}.

\section{Needs Pattern}
\label{sec:needs}

\begin{figure}[ht]
    \centerline{\includegraphics[width=0.6\textwidth]{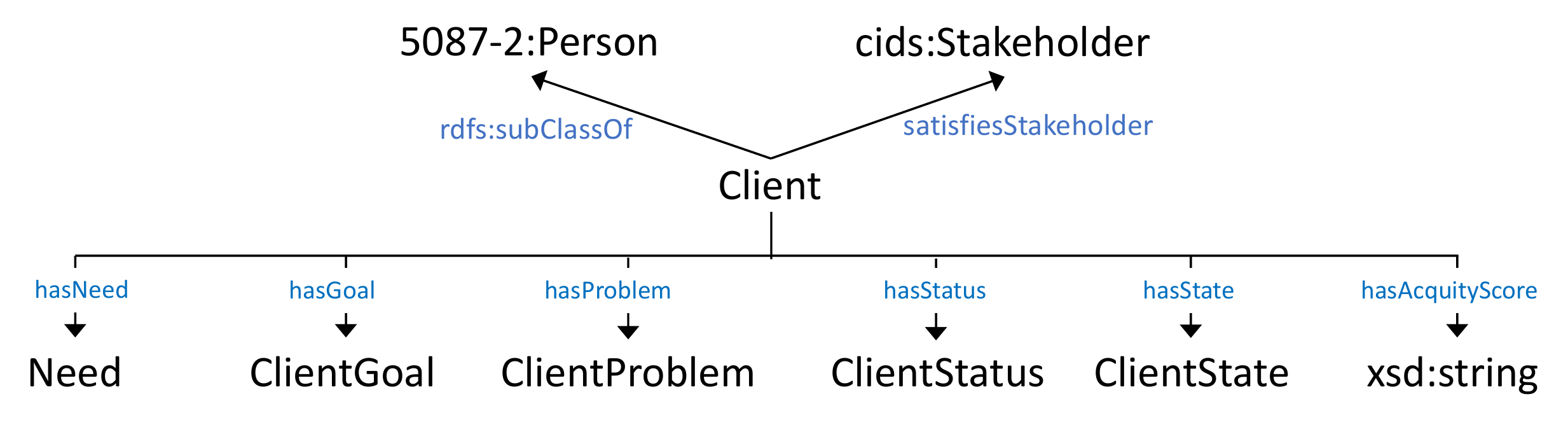}}
    \caption{\textit{Client} related classes for needs and outcomes.}
    \label{fig:client_needs}
\end{figure}

Needs provide the conceptual focus for service provision, and social work practice in general (\cite{gough2020, desmet2020, henwood2015}). Our representation of client needs, embodied in the Compass class \emph{ClientNeed}, defines the changes needed in a client's state. It relies on two elements: 1) a type of measurable personal feature that would be changed or maintained, such as employment or mental health state (pay rent, enroll kids in school), and 2) an action (e.g., \emph{improve, acquire, develop}). Based on the type of change sought, client needs can be divided into several categories, such as \emph{acquiring}-needs and \emph{improving}-needs. The operationalization of client needs as differences between the actual and desired (or prior- and post-service provision) values of measurable features allows us to infer a client’s current needs as well as check the satisfaction of those needs via observable client outcomes. The Compass \textit{ClientNeed} class is connected to the Compass \emph{ClientGoal} class, and through that, to the Compass \emph{ClientProblem} class. 

The \textit{ClientProblem} class is a cognitive representation of the discrepancy between an a client's actual and desired states. The use of the term \emph{problem} has a long pedigree in Social Work. \emph{Problems} are used to identify the key areas related to clients’ circumstances that are relevant for social work practice and are often recorded in client files as proxies for needs. The \textit{ClientGoal} class is a cognitive representation of reducing the gap between the desired and the actual state, if the two are different, or maintaining or ensuring a 0-distance between the desired and the actual state, if the two are the same. For example, a person who is ``couch surfing'' may wish to have a permanent place to live. Here, experiencing homelessness rather than being ``housed'' is a \emph{problem} that motivates the \emph{goal} of being stably housed. Given their current state and their stated goal, this person's need is \emph{to improve their housing}. 

As shown in Figure \ref{fig:client_needs}, the \textit{Client} class has properties that allow for clients to be connected with their states, problems, goals and needs. \textit{hasClientState} links to instances of \textit{ClientState} that specify the state the client is in. \textit{hasProblem} links to instances of \textit{ClientProblem} that specify the problems of the client. \textit{hasGoal} links to instances of \textit{ClientGoal} that specify the goals of the client. \textit{hasNeed} links to needs while \textit{hasAcquityScore} specifies their acuity level, as assessed by a healthcare professional, and recorded as a string value such as \{``Low'', ``Medium'', ``High''\} or \{``1'', ``2'', ``3''\}.  \textit{hasStatus} links to instances of Status that specify the status of the client. 

Client needs are met via need satisfiers (\cite{zins2001, human2018}), represented in Compass by the \emph{NeedSatisfier} class. The ``couch surfer'' in our example could have their need met by satisfiers such as \emph{rent supplements} and \emph{supportive short-term accommodations}. The connection between \textit{ClientNeed} and \textit{NeedSatisfier} is captured by the property \emph{hasNeedSatisfier}. A satisfier's type, such as \emph{pseudo satisfier} and \emph{inhibitor}, is specified via property \emph{hasType}. Property \emph{forNeed} connnects a need satisfier with needs it can fully or partially fulfill. The client states impacted by its provision are identified using property \emph{changes}. 



\section{Event Pattern}
\label{sec:event}

An \textit{Event} describes something that occurs at some location, at some time, and involves a \textit{Stakeholder}. It can describe the stakeholder as being the subject of an action, or a change of state. The basic \textit{Event} class has the following properties: \textit{occursAt} is the time interval during which the event occurs; \textit{hasLocation} is the placename where the event occurred; and \textit{previousEvent} links this event to a previous, related event while \textit{nextEvent} links it to the next related event, if any. Next, we provide extensions to basic \textit{Event} class.

\subsection{Client Event Pattern}
The \textit{ClientEvent} class, a subclass of \textit{Event}, is used to log significant events that are relevant to a client. They capture personal events, such as marriage, illness, employment, homelessness, and so on. \textit{ClientEvent} is a subclass of \textit{Event} and inherits all of its properties. It also adds the \textit{forClient} property. For example, one or more housing events may be used to determine that a client is homeless. A ``has'' property is defined for each event, for example, the property \textit{hasHousingEvent} identifies a \textit{HousingEvent} instance for the client, and allows us to track a client's transition from ``housed'' to ``institutionalized'', ``homeless'' and ``housed'' again. Events can be viewed as records of a client’s life, hence can be used retrospectively to understand a client's pathways. A subset of client events are listed next.
Additional client events related to a client property or status include \textit{EducationEvent, EmploymentEvent, MedicalEvent, MilitaryEvent, ImmigrationEvent, HousingEvent, NameEvent, GenderEvent, BirthEvent, DeathEvent, MaritalEvent, HomelessEvent}, and \textit{JusticeSystemEvent}.

The \textit{StakeholderEvent} captures events relevant to any stakeholder, not just a client. It extends the \textit{Event} class with the \textit{forStakeholder} property that identifies the \textit{cids:Stakeholder} associated with this Event instance.

\textit{ServiceFailureEvent} represents an event triggered when a barrier exists that prevents clients from using a service they may be otherwise eligible for. It extends the \textit{ClientEvent} class with the following properties. \textit{forService} identifies the Service or Activity this failure event indicates cannot be used by a client. \textit{hasCharacteristic} identifies the characteristic that caused the failure, such as an eligibility requirement that was not met (e.g. a person without identification cannot access their medication as it requires a health card). \textit{hasFailureType} identifies the \textit{cp:Service} or \textit{cids:Activity} type as a precondition for the service that might remove the barrier. For example, an ID clinic could provide the client with the required health card.

\subsection{Service Event Pattern}

The Compass \textit{Service} coverage competency questions also require the definition of events. A \textit{ServiceEvent} is an event that changes a client in some way. It extends the \textit{Event} class with the following properties. \textit{hasStatus} is the status of the service, and can take the value of \{\textit{scheduled, inProgress}, or \textit{completed}\}. \textit{atOrganization} identifies the organization providing the service. Finally, \textit{hasReferral} identifies the referral that led to the service event, if any.

For clients interacting with an instance of the \textit{Application} class, their activities are logged as \textit{ApplicationEvent} instances. It captures any information relevant to that event, including functional data and metadata. The \textit{hasApplication} property identifies the application this event was created in. \textit{hasUserStakeholder} identifies the stakeholder that was using the application when the event was created. \textit{hasSource} identifies the URL or unique address where this event originated, and may differ from the address listed for the \textit{Application} instance. \textit{hasMetaData} \textit{identifies} the information stored with the event.

\section{Evaluation}
\label{sec:eval}

The Compass extension to the Common Impact Data Standard, identified by the namespace prefix \textit{cp:}, was evaluated using a subset of competency questions listed in sections \ref{subsec:client-intro}, \ref{subsec:service-intro}, and \ref{subsec:outcome-intro}, and translated into SPARQL. The queries are run on a Compass-based knowledge graph and populated with synthetic data modeled after real data\footnote{Real data provided by Help Seeker Technologies: \url{helpseeker.org}.}. 


\noindent
\textbf{[Client Q-3]} \textit{Which services match my needs?}  The SPARQL query provided below retrieves services for a client with the internal identifier \textbf{cp:Client16} who is experiencing homelessness and struggles with addictions. Their needs are to reduce their suffering from addiction and improve their housing situation.

\lstset{language=sparql}
\begin{lstlisting}
SELECT DISTINCT ?service ?code WHERE {  
    BIND(cp:Client16 AS ?client).
    ?client  cp:hasNeed  ?need. 
    ?need  rdf:type  cp:ClientNeed.
    ?needSatisfier  rdf:type  cp:NeedSatisfier.
    ?need  cp:hasNeedSatisfier  ?needSatisfier.
    ?service  rdf:type  cp:Service ; cids:hasCode ?code ;
        cp:providesSatisfier ?needSatisfier.
}
\end{lstlisting}  

\noindent
Table \ref{tbl:client-need} lists the first 5 services retrieved by the sample query. Every service retrieved can address at least one of the client's needs via the satisfiers it provides, e.g., \emph{housing}, \emph{addiction treatment}, \emph{counseling}.   

\begin{table}[ht]
\centering
\caption{Client Need Q-3 SPARQL Query Results}
\label{tbl:client-need}
\begin{tabularx}{0.7\textwidth}{|X|X|}
\hline
\textbf{service}             & \textbf{code}                \\ \hline
cp:S17-Female-Shelter	     & cp:INST-Temporary\_Shelter   \\ \hline	
cp:S10-1-Shelter  	         & cp:INST-Shelter	            \\ \hline
cp:S14-Housing-For-Homeless  & cp:INST-Housing              \\ \hline	
cp:S15-A0-Addiction-Services &	cp:INST-Addiction\_Services \\ \hline 
cp:S06-1-Counseling	         & cp:INST-Counseling           \\ \hline 
\end{tabularx}
\end{table}


\noindent
\textbf{[Client CQ-6]} \textit{If I don't like the program what are my other options?} This question captures the situation where the initial service proposed to the client was rejected by them, and other service must be found that provides satifiers appropriate for their needs. The sample query assumes that the client is a homeless female residing in Area0, whose need is to improve her housing situation.


\lstset{language=sparql}
\begin{lstlisting}    
SELECT DISTINCT ?service ?code WHERE {  
    BIND(cp:NS-Housing AS ?needSatisfier).
    BIND(cp:Comp-Inst-Female-Homeless-Area0 AS ?compChar)
    ?service  rdf:type  cp:Service ;
        cids:hasCode ?code ; cp:hasRequirement ?compChar ;
        cp:providesSatisfier  ?needSatisfier.
}
\end{lstlisting}     



\noindent
Table \ref{tbl:need-satisfiers} lists the services retrieved by the sample query. Every service retrieved provides housing and has as the eligibility conditions of being female, experiencing homelessness and residing in ``Area0.''

\begin{table}[ht]
\centering
\caption{Need Satisfiers-based Client Q-6 SPARQL Query Results}
\label{tbl:need-satisfiers}
\begin{tabularx}{0.7\textwidth}{|X|X|}
\hline
\textbf{service}   &  \textbf{code}          \\ \hline
cp:S17-Female-Shelter & cp:INST-Temporary\_Shelter \\ \hline
cp:S10-1-Shelter      & cp:INST-Shelter           \\ \hline
\end{tabularx}
\end{table}

\noindent
\textbf{[Client Q-7.1]} \textit{What are you going to do with my data in the program?} This question requires listing of requirements for a particular service, namely \textit{cp:S06-1-Counseling}, limited to codes instances of class  \textit{cp:CL-Info\_Privacy}, that defines for data-privacy codes. 

    \lstset{language=sparql}
    \begin{lstlisting}
SELECT DISTINCT ?service ?dataReq WHERE { 
    BIND(cp:S06-1-Counseling AS ?service).
    ?dataReq rdf:type cp:CL-Info_Privacy.
    {?service cp:hasRequirement [cids:hasCode ?dataReq]. 
    } UNION {
        ?service cp:hasRequirement [
            rdf:type cids:CompositeCharacteristic   ; 
            oep:hasPart [cids:hasCode ?dataReq]]. }}   \end{lstlisting}     

\noindent
In Table \ref{tbl:q7-1} we see that requirements related to information privacy, namely \textit{cp:INST-Doctor\_Yes} stating that a medical doctor needs access to the client's data, and
\textit{cp:INST-Service\_Used\_Yes} stating the service itself needs access to the client's data.

\begin{table}[ht]
\centering
\caption{Client Q-7.1 SPARQL Query Results}
\label{tbl:q7-1}
\begin{tabularx}{0.7\textwidth}{|X|X|}
\hline
\textbf{service}      & \textbf{dataReq}             \\ \hline
cp:S06-1-Counseling & cp:INST-Doctor\_Yes        \\ \hline
cp:S06-1-Counseling & cp:INST-Service\_Used\_Yes \\ \hline
\end{tabularx}
\end{table}




\noindent
\textbf{[Service Q-2]} \textit{How long did client \#2 stay in counselling?} This question queries \textit{cp:ServiceEvent} instances for entries related to client \textit{cp:Client2} accessing counseling services, identified by the service taxonomy code \textit{cp:INST-Counseling}. For each record found, it calculates the number of weeks between the data properties \textit{time:hasBeginning} and \textit{time:hasEnd}, indicating when the counseling services began and finished\footnote{Date arithmetic functions for \textit{ofn:} and \textit{spif:} defined at \url{https://graphdb.ontotext.com/documentation/9.10/free/sparql-functions-reference.html}.}. Finally, the sum of weeks is displayed in the \textbf{weeks} column of Table \ref{tbl:service-q2}

    \lstset{language=sparql}
    \begin{lstlisting}
SELECT DISTINCT ?client ?weeks WHERE {  
    BIND(cp:Client2 AS ?client).
    ?serviceEvent rdf:type cp:ServiceEvent ; 
        cp:forClient ?client ;
        time:hasBeginning ?beg; time:hasEnd ?end;
        cids:hasCode cp:INST-Counseling.
    BIND((ofn:weeksBetween(
        spif:parseDate(?end, "yyyy-MM-dd'T'HH:mm:ss.SSS"),   
        spif:parseDate(?beg, "yyyy-MM-dd'T'HH:mm:ss.SSS"))) 
        AS ?weeks). }  
\end{lstlisting}

\begin{table}[ht]
\centering
\caption{Service Q-2 SPARQL Query Results}
\label{tbl:service-q2}
\begin{tabularx}{0.48\textwidth}{|X|X|}
\hline
\textbf{client} & \textbf{weeks} \\ \hline
cp:Client2    & 43             \\ \hline
\end{tabularx}
\end{table}


\noindent
\textbf{[Outcome Q-1]} \textit{What are the associated priority demographic groups?} This question asks about the service usage of specific demographics, and to have them ordered by priority, where priority here means the largest population impacted. The query begins by identifying all clients that use services, as defined by the \textit{cp:ServiceEvent} class in the knowledge graph. The results are accumulated by the stakeholder (\emph{?sh}) each client satisfies. Here, the stakeholder is defined by characteristics defined by client taxonomy codes and the location they reside.

    \lstset{language=sparql}
    \begin{lstlisting}
SELECT ?loc ?sh (COUNT(?sh) AS ?count) WHERE { 
    ?serviceEvent rdf:type cp:ServiceEvent ;
                  cp:forClient [cp:satisfiesStakeholder ?sh].
    ?sh a cids:Stakeholder ;
                 i72:located_in ?loc.
    { ?sh cids:hasCharacteristic [cids:hasCode ?demo].
    } UNION {
        ?sh cids:hasCharacteristic 
            [ a cids:CompositeCharacteristic;
                oep:hasPart [cids:hasCode ?demo]]
}} GROUP BY ?sh ?loc 
ORDER BY DESC(?count)
\end{lstlisting}     

\noindent
In Table \ref{tbl:outcome-q1} we see a demonstrative sample of results summed by location (\textbf{loc}), stakeholder (\textbf{sh}), and \textbf{code} identifying the service type. The \textbf{count} columns displays the number of clients that satisfy a particular stakeholder definition at the location specified. For example, we see that 18 female, housed youth clients in location ``Area0'' use counseling services.

\begin{table}[ht]
\centering
\caption{Outcome Q-1 SPARQL Query Results}
\label{tbl:outcome-q1}
\begin{tabularx}{\textwidth}{|p{80pt}|X|X|p{18pt}|}
\hline
\textbf{loc}         & \textbf{sh}                     & \textbf{code}                 & \textbf{count} \\ \hline
cp:Area0\_ Location & cp:sh-Female-Housed-Youth-in\_Area0    & cp:INST-Counseling          & 18             \\ \hline
cp:Area0\_ Location & cp:sh-Male-Youth-Addicted-in\_Area0    & cp:INST-Addiction\_Services & 15             \\ \hline
cp:Area0\_ Location & cp:sh-Female-Adult-Addicted-in\_Area0  & cp:INST-Addiction\_Services & 9              \\ \hline
cp:Area0\_ Location & cp:sh-Homeless-Male-Youth-in\_Area0    & cp:INST-Housing             & 6              \\ \hline
\end{tabularx}
\end{table}

\section{Discussion and Conclusion}
\label{sec:conclusion}

This paper introduces the Compass ontology and evaluates its ability to answer coverage-related competency questions.  The answers to the selected queries were extracted from a knowledge graph based on the Compass ontology and populated with client, service, service usage, and service funding data. The questions discussed in the evaluation section were selected from a list compiled by subject matter experts on behalf of various social services stakeholders, including service clients, service providers, and those interested in long-term outcomes, such as service managers and funders. 

Our evaluation demonstrates how coverage-related competency questions can be answered using the Compass ontology. \textbf{Client-coverage queries} identify clients that can use a service by listing requirements clients need to meet, the barriers they may face in meeting them, and offer alternatives when barriers cannot be overcome. \textbf{Service-coverage queries} successfully provide the cross-product service categories and targeted client populations. The queries also quantify resource usage and service availability over time. Such queries can provide information about specific services and clients or aggregate over categories of services and cohorts of clients based on client demographics, needs, and locations. Finally, the \textbf{outcome-coverage queries} answer questions similar to those about clients and services, but focus on the broader objectives of managerial and contributing stakeholders. These include obtaining quantitative information about the coverage of needs and highlighting potential gaps in service provisioning. These stakeholders can track their resources through the service-provisioning process across service categories, client cohorts, and locations.

Work continues on extending the client, service, and need taxonomies to ensure ontological concepts defined by Compass are shared and compared between organizations working towards similar or overlapping outcomes.

\bibliographystyle{unsrtnat}
\bibliography{library, need_satisfiers}

\end{document}